\documentclass[sigconf,natbib=false]{acmart}

\setcopyright{acmlicensed}
\copyrightyear{2026}
\acmYear{2026}
\acmDOI{XXXXXXX.XXXXXXX}
\acmConference[XXX]{ACM Conference}{2026}{XXX, XXX}
\acmISBN{978-1-4503-XXXX-X/2026/03}

\usepackage[ruled,lined,linesnumbered]{algorithm2e}
\usepackage{colortbl}
\usepackage{arydshln}
\usepackage{fontawesome5}
\usepackage{enumitem}
\usepackage{tabularx}
\usepackage{ragged2e}
\usepackage{multirow}

\usepackage{xspace}
\newcommand{\system}{\textsc{ViviDoc}\xspace}

\definecolor{bestblue}{RGB}{220, 235, 250}
\definecolor{secondgreen}{RGB}{235, 245, 230}
\definecolor{bestyellow}{RGB}{255, 243, 205}

\usepackage{pifont}

\title{\textsc{ViviDoc}: Generating Interactive Documents through Human-Agent Collaboration}

\author{Yinghao Tang}
\affiliation{%
  \institution{State Key Lab of CAD\&CG, Zhejiang University}
  \city{Hangzhou}
  \country{China}}
\email{tangyinghao@zju.edu.cn}

\author{Yupeng Xie}
\affiliation{%
  \institution{HKUST(GZ)}
  \city{Guangzhou}
  \country{China}}
\email{yxie686@connect.hkust-gz.edu.cn}

\author{Yingchaojie Feng}
\affiliation{%
  \institution{National University of Singapore}
  \city{Singapore}
  \country{Singapore}}
\email{yingchaojie@nus.edu.sg}

\author{Tingfeng Lan}
\affiliation{%
  \institution{University of Virginia}
  \city{Charlottesville}
  \state{VA}
  \country{USA}}
\email{tingfeng@virginia.edu}

\author{Jiale Lao}
\affiliation{%
  \institution{Cornell University}
  \city{Ithaca}
  \state{NY}
  \country{USA}}
\email{jiale@cs.cornell.edu}

\author{Yue Cheng}
\affiliation{%
  \institution{University of Virginia}
  \city{Charlottesville}
  \state{VA}
  \country{USA}}
\email{mrz7dp@virginia.edu}

\author{Wei Chen}
\affiliation{%
  \institution{State Key Lab of CAD\&CG, Zhejiang University}
  \city{Hangzhou}
  \country{China}}
\email{chenwei@zju.edu.cn}

\renewcommand{\shortauthors}{Tang et al.}

\begin{document}

\begin{abstract}
Interactive documents help readers engage with complex ideas through dynamic visualization, interactive animations, and exploratory interfaces. 
However, creating such documents remains costly, as it requires both domain expertise and web development skills. Recent Large Language Model (LLM)-based agents can automate content creation, but directly applying them to interactive document generation often produces outputs that are difficult to control.
To address this, we present \system, to the best of our knowledge the first work to systematically address  interactive document generation. \system introduces a multi-agent pipeline (Planner, Styler, Executor, Evaluator). To make the generation process controllable, we provide three levels of human control: (1) the Document Specification (DocSpec) with SRTC Interaction Specifications (State, Render, Transition, Constraint) for structured planning, (2) a content-aware Style Palette for customizing writing and interaction styles, and (3) chat-based editing for iterative refinement. We also construct \textsc{ViviBench}, a benchmark of 101 topics derived from real-world interactive documents across 11 domains, along with a taxonomy of 8 interaction types and a 4-dimensional automated evaluation framework validated against human ratings (Pearson $r > 0.84$). Experiments show that \system achieves the highest content richness and interaction quality in both automated and human evaluation. A 12-person user study confirms that the system is easy to use, provides effective control over the generation process, and produces documents that satisfy users. Our project homepage is available at \url{https://vividoc-homepage.vercel.app/}.
\end{abstract}



\maketitle
\section{Introduction}

\begin{figure*}[t]
    \centering
    \includegraphics[width=0.9\textwidth]{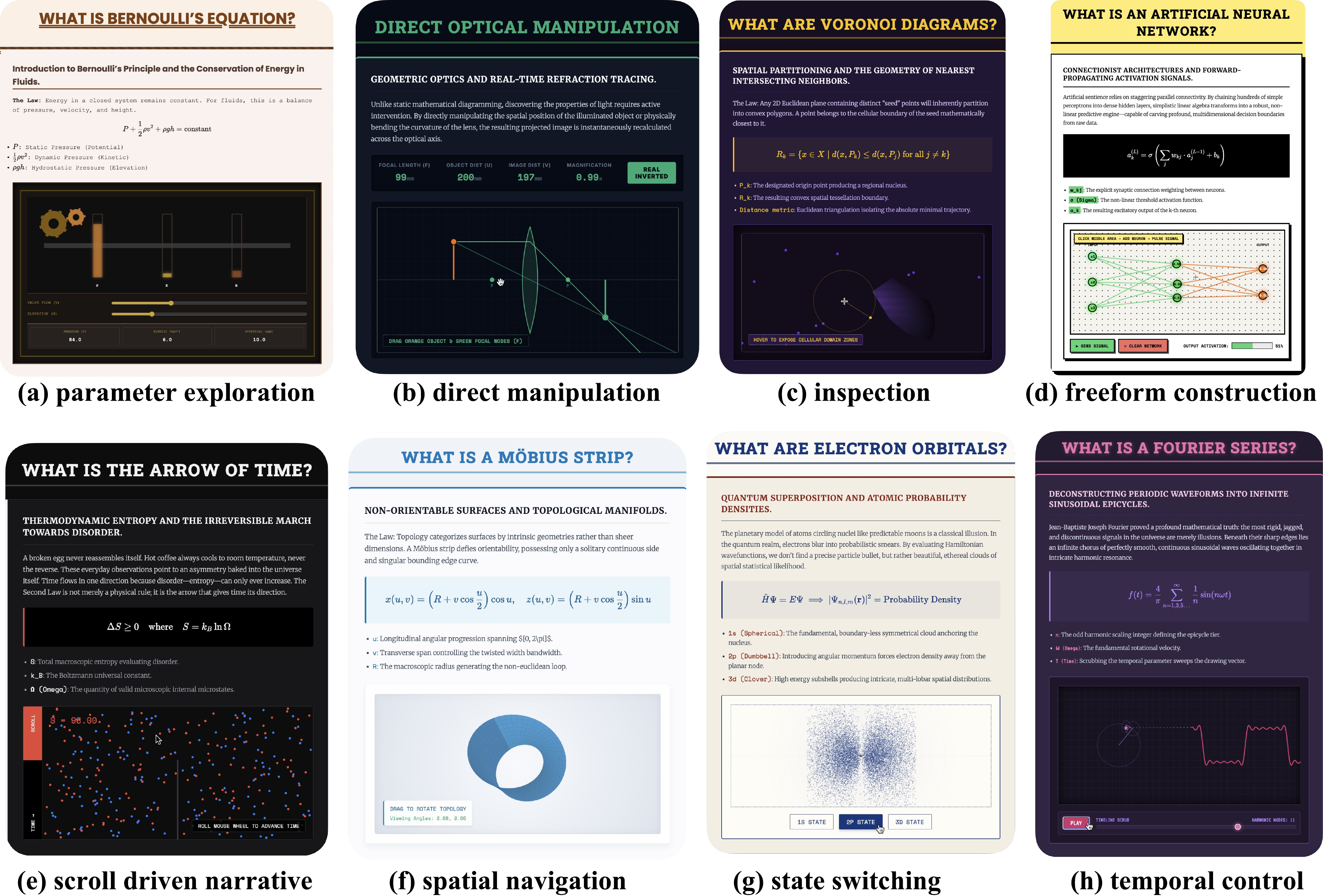}
    \caption{Eight interactive visualization examples generated by \system, covering all 8 interaction types in our taxonomy, with styles automatically adapted to each topic’s content.
        \textbf{(a) Parameter Exploration}: sliders adjust  flow rate and elevation parameters in real time.
        \textbf{(b) Direct Manipulation}: drag the object or focal points to update the lens equation live.
        \textbf{(c) Inspection}: hover to reveal a Voronoi cell and nearest-neighbor envelope.
        \textbf{(d) Freeform Construction}: click to place neurons and trigger animated signal propagation.
        \textbf{(e) Scroll-driven Narrative}: scroll to mix two particle gases and watch entropy rise.
        \textbf{(f) Spatial Navigation}: drag to rotate a 3D Möbius strip freely in space.
        \textbf{(g) State Switching}: switch quantum orbital states to redraw the electron probability cloud.
        \textbf{(h) Temporal Control}: play/pause and tune harmonics to build a Fourier series.
    }
    \label{fig:cases}
\end{figure*}

Interactive documents are an emerging communication medium that leverages the dynamic capabilities of the web to help readers engage with complex ideas~\cite{victor2011explorable}. Through interactive elements such as sliders, dropdowns, and direct manipulation controls, readers can actively explore concepts, observe cause-and-effect relationships, and build intuition through experimentation. This form of communication has been adopted across diverse domains including education~\cite{chi2014icap}, data-driven journalism~\cite{branch2012snow}, and scientific publishing~\cite{hohman2020communicating}. Studies show that interactive documents improve reader engagement and learning compared to static alternatives~\cite{hohman2020communicating}.

Despite their promise, creating interactive documents remains prohibitively costly. Producing a single piece, such as those on Explorable Explanations \cite{explorable_explanations} or Distill.pub~\cite{hohman2020communicating}, requires expertise in both domain knowledge and web development, often demanding days or weeks of effort per article. This bottleneck severely limits the production of high-quality interactive content.

In this paper, we aim to address this by leveraging the reasoning and code generation capabilities of LLM-based agents. To the best of our knowledge, this is the first work to systematically address interactive document generation. Although end-to-end LLM-based agents have achieved promising results in content creation tasks such as video generation~\cite{chen2025code2video}, slide design~\cite{zheng2025pptagent, liang2025slidegen}, and infographic production~\cite{tang2026igenbench}, naively applying them to interactive documents generation faces three key limitations.
First, the generation process is \textbf{largely uncontrollable}: there is a fundamental gap between authoring intent and the executable code that realizes it, and the agent resolves this gap based on its own implicit preferences.
Second, humans cannot effectively participate in the process: the agent operates as a \textbf{black box}, leaving educators no meaningful way to review or adjust intermediate decisions that shape the final output.
Third, \textbf{evaluation resources are scarce}: no grounded dataset or systematic evaluation framework exists for this task, making it difficult to compare approaches.

To address these limitations, we present \system, a human-agent collaborative system for generating interactive documents. \system is built on a multi-agent pipeline consisting of a Planner, Styler, Executor, and Evaluator. To make the generation process controllable and transparent, we provide three levels of human control:
(1) The Document Specification (DocSpec), a structured intermediate representation that organizes a document into knowledge units, each containing a text description for content generation and an Interaction Specification for visualization generation. The Interaction Specification uses a four-component decomposition of State, Render, Transition, and Constraint (SRTC), inspired by interactive visualization theory~\cite{munzner2014visualization}. SRTC can express all common interaction types found in real-world educational documents. Educators can review, modify, and refine both the content plan and the interaction design before any code is produced.
(2) A Style Palette, where the LLM analyzes the spec content and generates style options for users to customize the writing and interaction style of the document.
(3) Chat-based editing, which allows users to refine both the spec and the generated document through natural language conversation.

To enable systematic evaluation, we construct \textsc{ViviBench}, a benchmark grounded in 101 real-world interactive documents collected from over 60 websites across 11 domains. We reverse-engineer teaching topics from these documents, ensuring that evaluation is grounded in topics for which high-quality interactive content already exists. From 482 interaction instances in these documents, we derive a taxonomy of 8 interaction types (e.g., Parameter Exploration, State Switching, Direct Manipulation). We also design a 4-dimensional automated evaluation framework combining rule-based checks (Interaction Functionality, Efficiency) with LLM-as-Judge assessment (Content Richness, Interaction Quality), and validate its reliability through human evaluation alignment (Pearson $r > 0.84$).

We evaluate \system through both automated evaluation and human evaluation with 12 raters. Results show that \system achieves the highest content richness and interaction quality among all methods. A 12-person user study further confirms that the system is easy to use (5.0/5), that DocSpec, Style Palette, and chat editing provide effective control (all rated above 4.0/5), and that participants are satisfied with the generated documents (4.58/5).

We summarize our contributions as follows:
\begin{itemize}[noitemsep,leftmargin=10pt]
    \item We introduce the task of interactive document generation and identify three key challenges of naively applying LLM-based agents to this task: uncontrollability, absence of human participation, and lack of evaluation resources.

    \item We propose \system, a multi-agent system that provides three levels of human control: the Document Specification (DocSpec) with SRTC Interaction Specifications for structured planning, a content-aware Style Palette for style customization, and chat-based editing for iterative refinement, making the generation process controllable and transparent.

    \item We construct \textsc{ViviBench}, a benchmark grounded in 101 real-world interactive documents from over 60 websites across 11 domains. We derive a taxonomy of 8 interaction types from 482 interaction instances and design an automated evaluation framework validated against human ratings.

    \item We conduct comprehensive experiments including automated evaluation, human evaluation, and a user study, demonstrating that \system achieves the highest document quality and provides an effective and satisfying authoring experience.
\end{itemize}

\section{Related Work}
\label{sec:related} 

\subsection{Interactive Document}
Interactive documents, often conceptualized as explorable explanations~\cite{victor2011explorable}, are a transformative medium that leverages dynamic web capabilities to foster deep engagement with complex ideas. The theoretical foundation for such active reading environments traces back to early human-computer interaction paradigms, including Engelbart's framework for augmenting human intellect~\cite{engelbart2023augmenting} and foundational hypertext structures~\cite{nelson1965complex}, which were operationalized in early systems like PLATO~\cite{bitzer2007plato}. Today, empirical studies confirm that interactive documents significantly improve reader engagement and comprehension compared to static alternatives~\cite{chi2014icap, hohman2020communicating}. The medium has been successfully adopted across diverse fields, from immersive data journalism (e.g., The New York Times' Snow Fall~\cite{branch2012snow} and Bloomberg's climate visualizations~\cite{roston2015whats}) to scholarly communications exploring machine learning fairness~\cite{wattenberg2016attacking}.

Despite their efficacy, creating these artifacts remains prohibitively costly, demanding a rare intersection of deep domain knowledge and web development expertise. This bottleneck severely limits the availability of high-quality interactive content. To overcome these authoring barriers, we introduce \system, a human-agent collaborative system designed to systematically automate and control the generation of interactive documents.

\subsection{LLM Agents for Content Creation}

Recent advancements increasingly rely on LLMs and multi-agent systems to automate complex content creation by decomposing tasks across specialized agents \cite{lin2025creativity,zou2025llmbasedhumanagentcollaborationinteraction}. These frameworks have achieved success across various domains, including data visualization systems like  HAIChart~\cite{xie2024haichart}, VisPilot~\cite{wen2025exploring}, and DataLab~\cite{weng2025datalab}. Similarly, presentation agents like PPTAgent \cite{zheng2025pptagent}, and SlideGen \cite{liang2025slidegen} use structural schemas for coherent slide decks, while systems like LAVES \cite{yan2026beyond} extend multi-agent orchestration to generate synchronized educational videos.

However, applying such agentic approaches to interactive documents remains largely unexplored. Naive implementation faces critical limitations regarding uncontrollable generation lacking intent alignment, the black-box nature of agents that preclude meaningful human intervention, and the absence of specialized datasets and systematic evaluation methods. Addressing these fundamental bottlenecks, \system establishes a controllable, human-in-the-loop generation paradigm by structuring the collaborative process around an interpretable intermediate representation.

\begin{figure*}[t]
\centering
\includegraphics[width=0.9\textwidth]{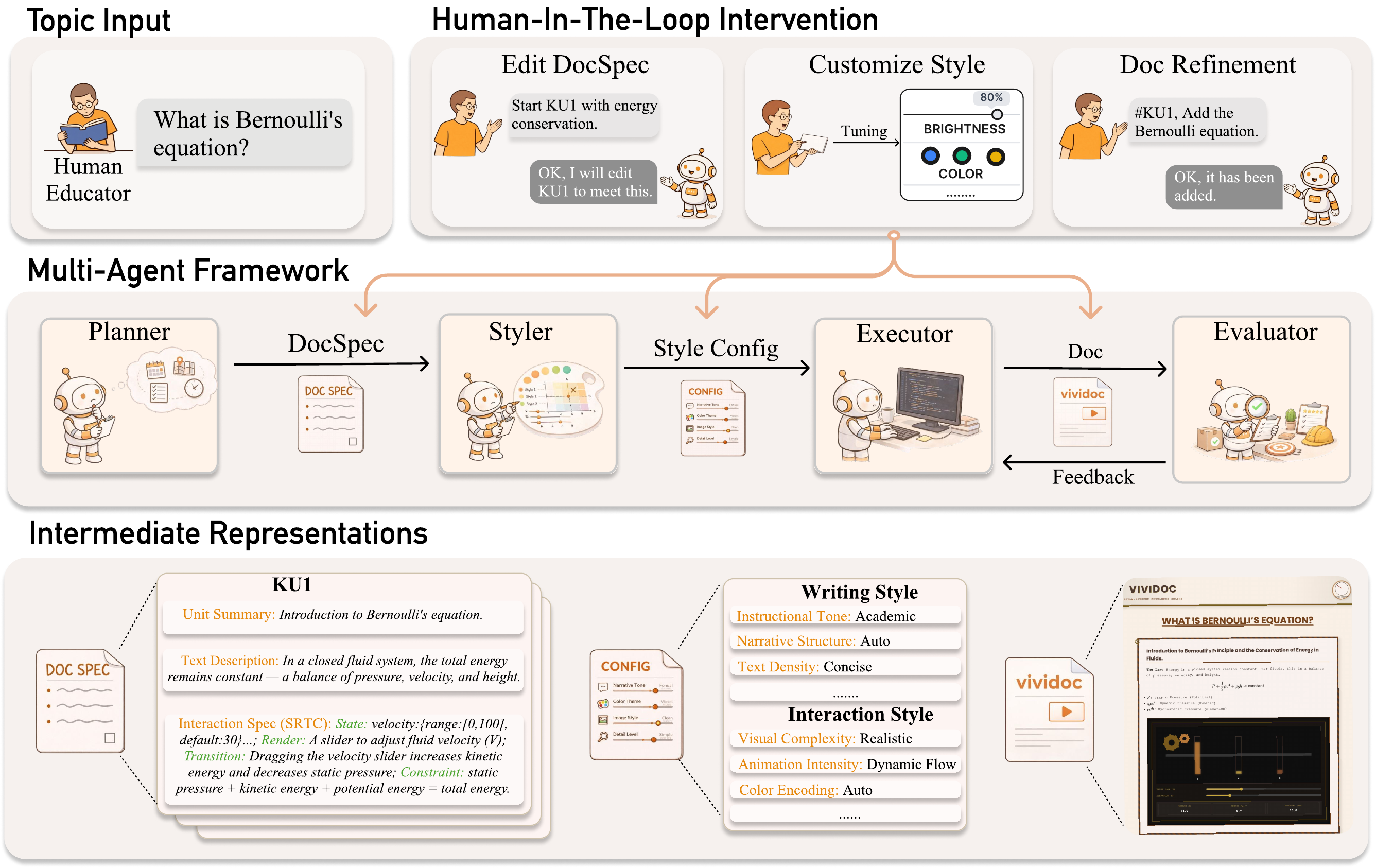}
\caption{The \system pipeline. Given a topic, the Planner generates a DocSpec consisting of knowledge units with text descriptions and SRTC Interaction Specifications. The Styler analyzes the DocSpec and generates a Style Palette for users to customize writing and interaction styles. The Executor generates the document code guided by the DocSpec and style instructions. The Evaluator checks the output for correctness. Users can intervene at three points: editing the DocSpec, customizing the Style Palette, and refining the document through chat.}
\label{fig:pipeline}
\end{figure*}

\section{System Overview}
\label{sec:system}
As illustrated in Figure~\ref{fig:pipeline}, \system generates interactive documents through a four-agent pipeline: Planner, Styler, Executor, and Evaluator, coordinated by a structured intermediate representation called the Document Specification (DocSpec). The pipeline exposes three points of human control: editing the DocSpec after planning, customizing style preferences through the Style Palette, and refining the generated document through chat-based editing.

\paragraph{Planner.} The Planner agent takes a topic (e.g., ``What is  Bernoulli's equation?'') and produces a DocSpec. It decomposes the topic into a sequence of knowledge units, each containing a text description that guides content generation and an SRTC Interaction Specification  that defines the interactive visualization. The Planner uses an LLM with structured output to ensure the DocSpec conforms to a predefined schema. We describe the DocSpec structure in detail in Section~\ref{sec:spec}.

\paragraph{Styler.} The Styler agent analyzes the DocSpec content and generates a Style Palette, a set of style dimensions with multiple options for the user to choose from. The dimensions are organized into two categories: writing style (e.g., narrative tone, terminology level) and interaction style (e.g., visual complexity, animation intensity). Each dimension offers 2--3 LLM-generated options along with an ``Auto'' option (delegating the choice to the LLM) and a ``Custom'' option (free-text input). The selected options are compiled into natural language instructions and injected into the Executor prompts: writing style instructions for Step 1 (text generation) and interaction style instructions for Step 2 (visualization generation).

\paragraph{Executor.} The Executor agent takes the DocSpec and style instructions and generates the final HTML document. It processes each knowledge unit in two steps. In Step 1, it generates the text content as an HTML fragment, guided by the text description, the writing style instructions, and the context of previously generated sections to maintain consistency. In Step 2, it generates the interactive visualization as HTML, CSS, and JavaScript, guided by the SRTC Interaction Specification and the interaction style instructions.

\paragraph{Evaluator.} The Evaluator agent checks the generated document for correctness. It validates the HTML structure, verifies that all knowledge units have been successfully generated, and checks that each section passes HTML validation. If issues are found, the Evaluator provides feedback that can trigger re-execution of specific components.

\paragraph{Human-in-the-Loop Control.} Human control spans three points in the pipeline. First, after the Planner produces the DocSpec, users can reorder knowledge units, modify text descriptions, adjust interaction parameters (e.g., variable ranges, control types), or add and remove units entirely. Because the DocSpec is structured rather than free-form, edits are targeted and predictable: changing a slider range in the Interaction Specification directly affects the generated visualization without requiring the user to write code. Second, the Style Palette allows users to customize the writing and interaction styles of the document by selecting from LLM-generated options or providing custom instructions. Third, after the Executor produces the document, users can refine the output through a chat-based interface, describing desired changes in natural language. The system parses these requests and applies corresponding edits to the document.

The DocSpec serves as a contract between pipeline stages: the Planner expresses intent in a structured form, the Styler translates user preferences into generation instructions, the Executor implements the plan as code, and the Evaluator verifies the result. This decomposition isolates the most error-prone step, translating intent into code, and constrains it with a structured specification rather than ambiguous natural language.

\section{Document Specification}
\label{sec:spec}

The Document Specification (DocSpec) is the structured intermediate representation at the core of \system. It bridges the gap between a high-level topic and the generated interactive document by decomposing the content into a sequence of knowledge units, each with explicit instructions for both text and interaction generation.

\subsection{Structure}

A DocSpec consists of a topic and an ordered list of knowledge units. Each knowledge unit contains three components:

\begin{itemize}[noitemsep,leftmargin=10pt]
    \item A \textbf{unit summary} that briefly states the concept covered.
    \item A \textbf{text description} that provides a self-contained guide for generating the explanatory text of the section. It specifies what the reader should understand after reading, without prescribing exact wording.
    \item An \textbf{Interaction Specification} that defines the interactive visualization using the SRTC decomposition described below.
\end{itemize}

The text description and Interaction Specification serve different steps of the Executor: the former guides Step 1 (text generation) and the latter guides Step 2 (visualization generation). This separation allows each step to operate independently with minimal ambiguity. The Interaction Specification supports all common interaction patterns such as parameter exploration, direct manipulation, and mode switching (see Figure \ref{fig:cases}). Due to space limitations, we provide the corresponding specifications for these cases in the supplementary material.

\subsection{Interaction Specification: SRTC}

Munzner's What-Why-How framework~\cite{munzner2014visualization} provides a foundational decomposition for visualization design: What data is being visualized, How it is visually encoded and interacted with, and Why the user engages with it. We adapt this framework to the setting of LLM-based generation by introducing the SRTC Interaction Specification, which decomposes each interactive visualization into four components:

\begin{itemize}[noitemsep,leftmargin=10pt]
    \item \textbf{State (S)}: The variables underlying the visualization, including their types, domains, default values, and derivation rules. Variables are either user-controllable (e.g., a slider with a specified range) or derived from other variables (e.g., a formula). This corresponds to Munzner's What.
    \item \textbf{Render (R)}: A description of how the state maps to visual elements on screen, such as geometric shapes, labels, or charts. This corresponds to the visual encoding aspect of How.
    \item \textbf{Transition (T)}: A description of how user actions modify the state, specifying the cause-and-effect relationship between input events and state changes. This corresponds to the interaction idiom aspect of How.
    \item \textbf{Constraint (C)}: The key invariant that the visualization is designed to demonstrate. This is the core insight the reader should discover through interaction. This corresponds to Munzner's Why, adapted from ``what task is the user performing'' to ``what should the reader observe.''
\end{itemize}

\paragraph{Example.} Table~\ref{tab:spec-example} contrasts a natural language interaction description with its corresponding SRTC specification for a visualization about $\pi$. The natural language version leaves implicit what variables exist, how they relate, what appears on screen, and what invariant the reader should notice. The SRTC version makes each of these explicit. The State defines a user-controllable variable $r$ (a slider over $[0.5, 5]$) and three derived variables: circumference $C = 2\pi r$, diameter $D = 2r$, and their ratio $C/D$. The Render specifies that the visualization displays a circle whose size reflects $r$, along with labels for $C$, $D$, and the ratio. The Transition links the slider interaction to the state: dragging the slider changes $r$, and all derived variables update automatically. Finally, the Constraint encodes the key invariant that the reader should discover: $C/D \approx 3.14159$ regardless of $r$.

\begin{table}[t]
  \caption{Comparison of a natural language interaction description and its SRTC Interaction Specification for a visualization about $\pi$.}
  \label{tab:spec-example}
  \small
  \centering
  \begin{tabularx}{\columnwidth}{@{}lX@{}}
\toprule
\multicolumn{2}{@{}l}{\textbf{Natural language description}} \\
\midrule
\multicolumn{2}{@{}p{\columnwidth}@{}}{``The reader can adjust the radius of a circle using a slider, and as the radius changes, the visualization updates the circle while recalculating and displaying the circumference, diameter, and ratio, allowing the reader to see that the ratio stays roughly the same.''} \\
\midrule
\multicolumn{2}{@{}l}{\textbf{SRTC Interaction Specification}} \\
\midrule
\textbf{S} & \texttt{r}: slider $[0.5, 5]$, default 1; \texttt{C}: derived $2\pi r$; \texttt{D}: derived $2r$; \texttt{ratio}: derived $C/D$ \\
\textbf{R} & A circle whose size reflects $r$; labels showing $C$, $D$, and ratio \\
\textbf{T} & Dragging the slider changes $r$; $C$, $D$, ratio update automatically \\
\textbf{C} & ratio $\approx 3.14159$ regardless of $r$ \\
\bottomrule
  \end{tabularx}
\end{table}

\section{Benchmark}
\label{sec:benchmark}

To systematically evaluate the quality of generated interactive documents, we construct \textsc{ViviBench}, a benchmark consisting of a curated topic dataset grounded in real-world interactive documents and an automated evaluation framework covering four complementary dimensions.

\subsection{Dataset}
\label{sec:dataset}

A key challenge in benchmarking interactive document generation is obtaining topics that reflect genuine needs rather than synthetic or contrived examples.
To address this, we adopt a reverse-engineering approach: we collect 101 real-world interactive documents from 63 distinct websites across 11 subject areas, including platforms such as \texttt{setosa.io}~\cite{setosa} and \texttt{distill.pub}~\cite{hohman2020communicating}.      
For each document, we use an LLM to extract the core topic as a concise natural-language description (e.g., ``Exponential growth as repeated multiplication'').
These extracted topics serve as the input prompts for all generation methods.
Due to space limitations, we provide the distribution of topics across subject areas in the supplementary material.

From these 101 documents, we also extract 482 interaction instances. Three visualization experts collaboratively classify these instances into 8 interaction types based on interaction intent such as Parameter Exploration and State Switching (see Figure \ref{fig:cases}). This taxonomy inspires the design of the SRTC Interaction Specification (Section~\ref{sec:spec}), which can express all 8 identified types.


\begin{figure*}[t]
  \centering
  \includegraphics[width=0.85\textwidth]{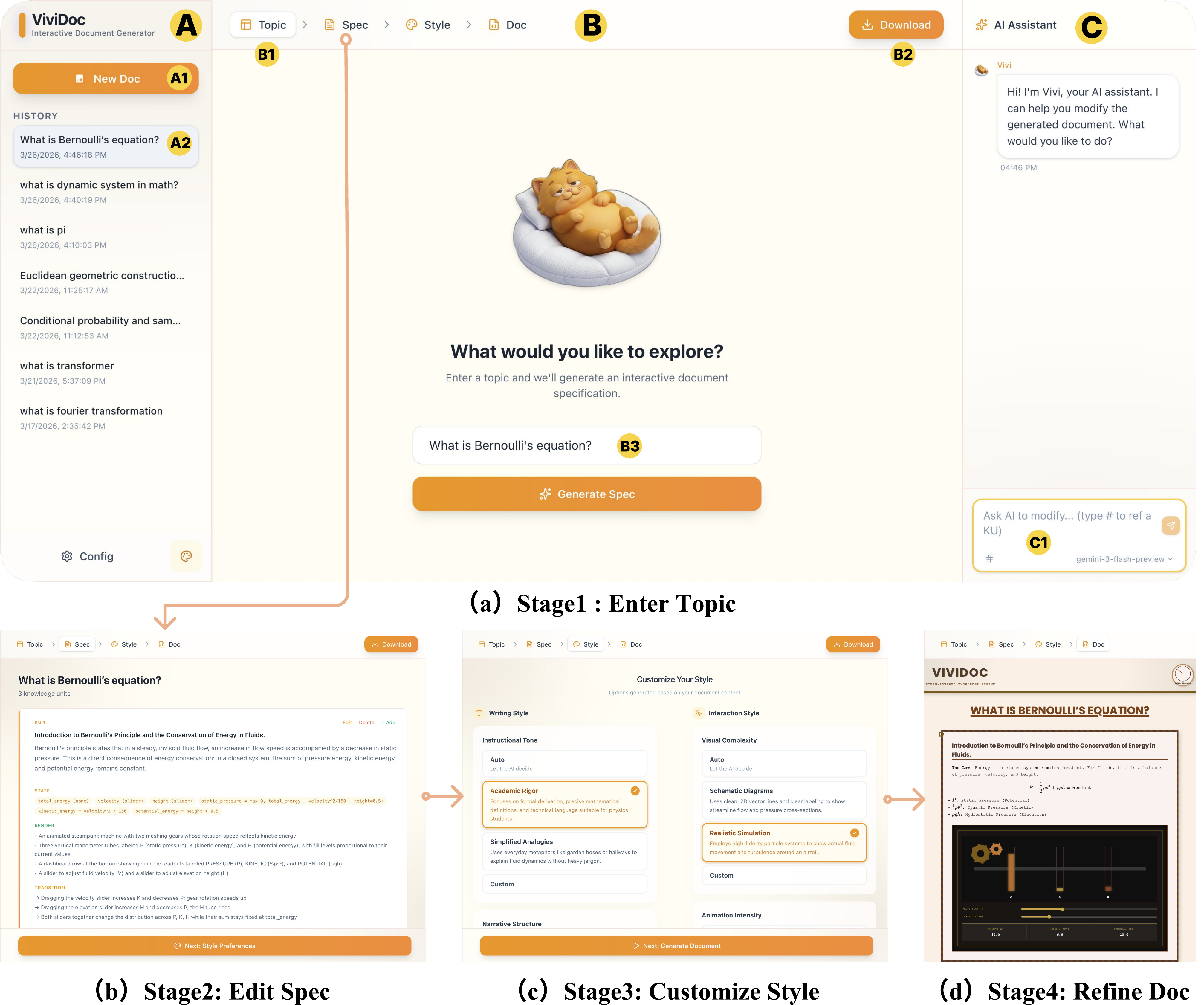}
  \caption{The \system user interface. Top: the main view with (A) sidebar for history and new document creation, (B) center panel with topic input and four-stage navigation bar, and (C) AI chat assistant. Bottom (left to right): the Spec stage showing editable knowledge units, the Style stage with writing and interaction style options, and the Doc stage displaying the generated interactive document.}
  \label{fig:system}
\end{figure*}

\subsection{Evaluation Metric}
\label{sec:evaluation}

We evaluate generated documents along four dimensions using a two-layer evaluation framework. The first layer applies deterministic, rule-based checks to verify functional correctness and generation efficiency. The second layer employs LLM-as-Judge to assess higher-level content and interaction quality~\cite{zheng2023judging, xie2025visjudge}.

\paragraph{Layer 1: Rule-based Evaluation.}
Two dimensions are evaluated through automated checks, providing reproducible and deterministic assessments:

\begin{itemize}[nosep,leftmargin=*]
  \item \textbf{Interaction Functionality (IF):} Tests whether interactive elements (buttons, sliders, checkboxes, dropdowns) respond to programmatic interaction using Playwright browser automation. For each element, we record the DOM state before and after triggering the interaction; elements that produce a DOM change are counted as functional. The score is the ratio of responsive elements to total interactive elements found.
  \item \textbf{Efficiency (Eff):} Measures the generation throughput as the ratio of output HTML length (in characters) to generation time (in seconds). This metric captures how efficiently each method utilizes its LLM calls to produce content.
\end{itemize}

\paragraph{Layer 2: LLM-as-Judge Evaluation.}
Two dimensions are assessed by prompting an LLM to score each document on a 1--5 Likert scale with detailed rubrics:

\begin{itemize}[nosep,leftmargin=*]
  \item \textbf{Content Richness (CR):} Whether the document covers the topic with sufficient breadth (e.g., multiple sub-concepts) and depth (e.g., explanations, examples, connections between ideas). We extract the main textual content from the HTML, stripping scripts and styling, so the judge focuses on the educational substance rather than code volume.
  \item \textbf{Interaction Quality (IQ):} A composite metric defined as $\text{IQ} = \text{ID} \times \text{IF}$, where ID (Interaction Design) is the LLM-judged score for whether interactive elements serve a clear purpose, and IF is the rule-based Interaction Functionality score. This formulation ensures that only functionally working interactions contribute to the quality score.
\end{itemize}

\begin{figure*}[htbp]
  \centering
  \includegraphics[width=0.85\linewidth]{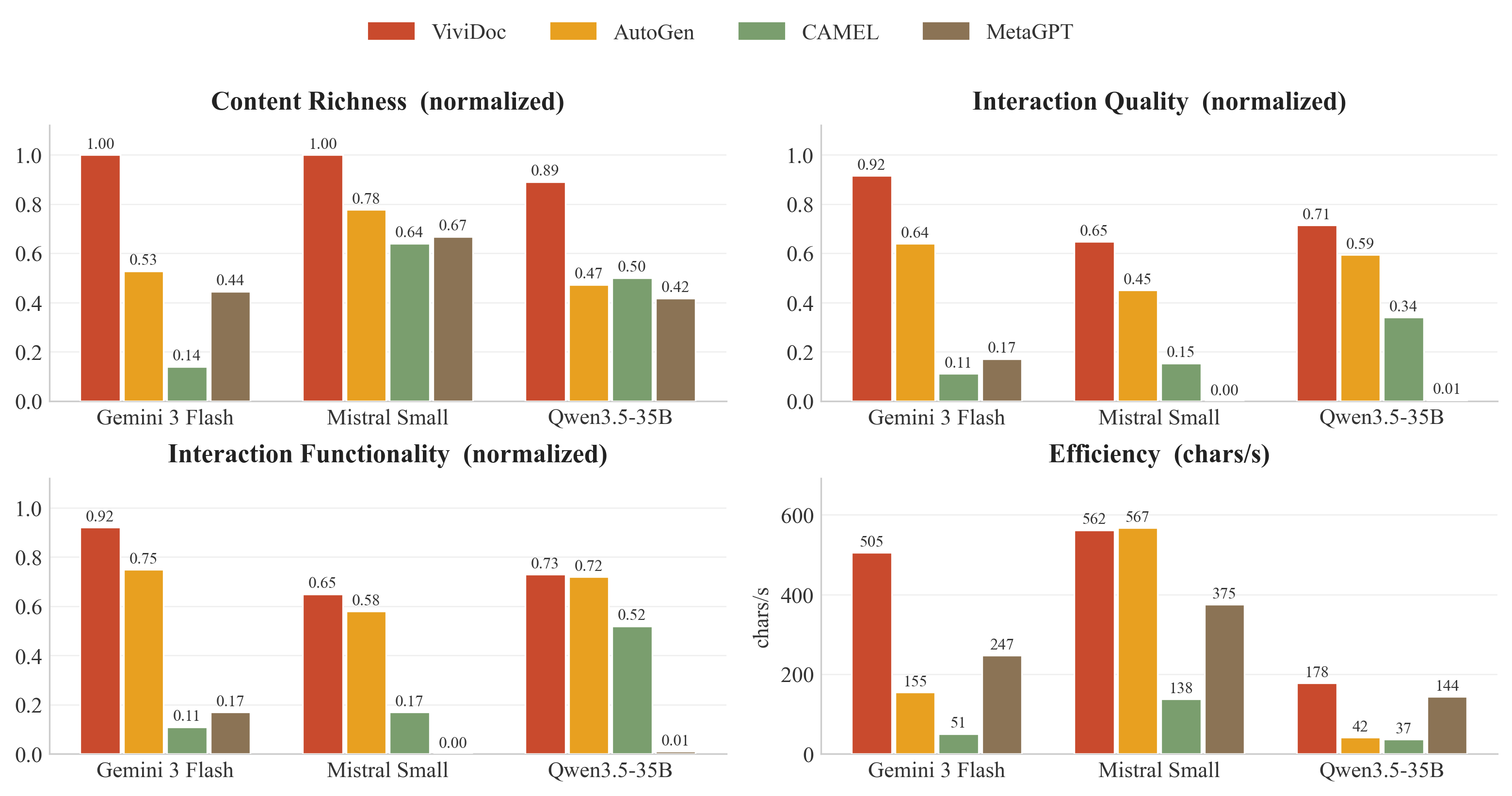}

  \caption{Automated evaluation results for \system vs.\ three multi-agent baselines across three backbone LLMs.
    Content Richness (CR) and Interaction Quality (IQ) are normalized to $[0,1]$.
    Interaction Functionality is on a 0--1 scale; Efficiency is measured in characters per second (chars/s).}
  \label{fig:auto_eval}
  \vspace{-1em}
\end{figure*}

\section{User Interface}
\label{sec:ui}

Figure~\ref{fig:system} shows the \system user interface, which consists of three panels: a sidebar (A), a center panel (B), and an AI assistant panel (C). The sidebar (A) provides a ``New Doc'' button (A1) to start a new generation and a history list (A2) of previously generated documents. The center panel (B) guides users through a four-stage workflow indicated by the top navigation bar (B1): Topic, Spec, Style, and Doc. A download button (B2) allows exporting the final document. The AI assistant panel (C) provides a chat interface (C1) for refining the spec or the generated document through natural language conversation.

\paragraph{Stage 1: Topic.} The user enters a topic of interest in the input field (B3) and clicks ``Generate Spec'' to start the pipeline.

\paragraph{Stage 2: Spec.} The Planner generates a DocSpec, displayed as a list of knowledge units (Figure \ref{fig:system}b). Each unit shows its title, text description, and SRTC Interaction Specification. Users can edit, reorder, add, or delete knowledge units before proceeding.

\paragraph{Stage 3: Style.} The Styler generates a Style Palette with two columns: Writing Style and Interaction Style (Figure \ref{fig:system}c). Each column contains several dimensions with LLM-generated options. Users can select an option, choose ``Auto'' to let the LLM decide, or provide custom instructions.

\paragraph{Stage 4: Doc.} The Executor generates the final interactive document, rendered in the center panel (Figure \ref{fig:system}d). Users can interact with and download the document. The chat assistant (C1) allows further refinement through natural language requests.

\section{Experiments}
\label{sec:experiment}

\subsection{Effectiveness of \system}
\label{sec:auto_eval}

\paragraph{Setup.}
We generated interactive documents for all topics in
\textsc{ViviBench} using \system and three multi-agent
baselines: AutoGen~\cite{wu2024autogen},
CAMEL~\cite{li2023camel}, and MetaGPT~\cite{hong2023metagpt}.
Each method was run with three backbone LLMs
(Gemini 3 Flash~\cite{team2023gemini},
Mistral Small~\cite{mistral2024small}, and
Qwen 3.5-35B~\cite{bai2023qwen}) to assess robustness
across model capabilities.
For fair comparison in the quantitative evaluations
(Sections~\ref{sec:auto_eval}--\ref{sec:human_eval}),
we remove human control from \system and operate its full
pipeline (Planner $\to$ Styler $\to$ Executor $\to$ Evaluator)
in fully automated mode with the style set to auto.
For the three baselines, we implement the equivalent roles
from our pipeline within their respective frameworks,
using the same topics as input.
All LLM-as-Judge evaluations use Gemini 3.1
Pro~\cite{team2023gemini} as the judge model.

\paragraph{Results.}
Figure~\ref{fig:auto_eval} reports the four metrics per
method-backbone combination.
\system consistently achieves the highest Content Richness
and Interaction Quality across all backbones, substantially
outperforming all baselines.
For example, with Gemini 3 Flash as the backbone, \system
achieves a normalized CR of 1.00 and IQ of 0.92, compared
to 0.53 and 0.64 for the strongest baseline AutoGen.
Notably, CAMEL and MetaGPT produce documents with near-zero IQ scores (< 0.05 across all backbones), indicating that their generated interactions rarely function correctly---the rule-based IF component effectively penalizes non-functional interactions.
Furthermore, \system achieves significantly higher end-to-end Efficiency. For instance, when using Gemini 3 Flash as the backbone, the efficiency of \system (505 chars/s) is 3.3$\times$, 9.9$\times$, and 2.0$\times$ higher than AutoGen, CAMEL, and MetaGPT, respectively. The above results demonstrate that \system, as a curated pipeline for the interactive document generation task, vastly outperforms general-purpose multi-agent frameworks in both document generation quality and efficiency.

\subsection{Effectiveness of DocSpec}
\label{sec:docspec_eval}

To isolate the contribution of DocSpec, we compare \system against a Naive Agent baseline that utilizes our generation pipeline but removes the DocSpec generated by the Planner as guidance, instead generating the interactive document directly end-to-end based on the topic.

\paragraph{Results.}
Table~\ref{tab:docspec} shows the comparison across all three backbones.
\system consistently outperforms Naive Agent on CR, IQ, and IF across all backbone LLMs, confirming that DocSpec's structured planning leads to richer content and more functional interactions.
The improvement is most pronounced on IQ (+41\% with Gemini 3 Flash), highlighting that generation without interaction specifications struggles to produce coherent interactive elements even when content quality is reasonable. In terms of efficiency,
\system introduces additional planning overhead, resulting
in modest efficiency reductions for Gemini and Mistral
(6\% and 19\%, respectively). Qwen exhibits a larger
drop (50\%), as the additional planning calls consume
a greater share of total time on a lower-throughput
model. Overall, the efficiency trade-off is well
justified by the consistent gains in content richness
and interaction quality.

\begin{table}[htbp]
  \caption{DocSpec effectiveness: \system vs.\ Naive Agent across three backbone LLMs. CR: Content Richness (1--5), IQ: Interaction Quality (0--5), IF: Interaction Functionality (0--1), Eff: Efficiency (chars/s). \colorbox{bestyellow}{Yellow}: best per backbone.}
  \label{tab:docspec}
  \centering
  \small
  \begin{tabular}{l l cccc}
    \toprule
    \textbf{Backbone} & \textbf{Method} & \textbf{CR}$\uparrow$ & \textbf{IQ}$\uparrow$ & \textbf{IF}$\uparrow$ & \textbf{Eff}$\uparrow$ \\
    \midrule
    \multirow{2}{*}{Gemini 3 Flash}
      & Naive Agent & 4.00 & 3.24 & 0.85 & \cellcolor{bestyellow}539 \\
      & \system     & \cellcolor{bestyellow}5.00 & \cellcolor{bestyellow}4.58 & \cellcolor{bestyellow}0.92 & 505 \\
    \midrule
    \multirow{2}{*}{Mistral Small}
      & Naive Agent & 4.44 & 2.36 & 0.61 & \cellcolor{bestyellow}695 \\
      & \system     & \cellcolor{bestyellow}5.00 & \cellcolor{bestyellow}3.24 & \cellcolor{bestyellow}0.65 & 562 \\
    \midrule
    \multirow{2}{*}{Qwen3.5-35B}
      & Naive Agent & 4.11 & 2.76 & 0.70 & \cellcolor{bestyellow}355 \\
      & \system     & \cellcolor{bestyellow}4.56 & \cellcolor{bestyellow}3.57 & \cellcolor{bestyellow}0.73 & 178 \\
    \bottomrule
  \end{tabular}
\end{table}

\subsection{Human Evaluation}
\label{sec:human_eval}

To validate the automated evaluation, we conducted a human evaluation study with 12 raters.

\paragraph{Setup.}
We selected 9 topics randomly from \textsc{ViviBench}, ensuring coverage across different subject areas, and generated documents using all four methods with the Gemini 3 Flash backbone, yielding 36 documents.
We recruited 12 human raters and divided them into 3 groups of 4, with each group evaluating 12 documents (3 topics $\times$ 4 methods) in a blind setting.
Raters scored each document on Content Richness (CR) and Interaction Design (ID) using a 1--5 Likert scale.

\paragraph{Human Scores.}
Table~\ref{tab:human_scores} shows the mean human scores per method.
\system achieves the highest scores on both dimensions (CR: 4.43, ID: 4.33), followed by AutoGen.
CAMEL receives the lowest scores, consistent with the automated evaluation results.

\begin{table}[htbp]
  \caption{Mean human evaluation scores by method (Gemini 3 Flash backbone), averaged across 12 raters and 9 topics. \colorbox{bestyellow}{Yellow}: best per metric.}
  \label{tab:human_scores}
  \centering
  \small
  \begin{tabular}{l cccc}
    \toprule
    & \textbf{AutoGen} & \textbf{MetaGPT} & \textbf{CAMEL} & \textbf{\system} \\
    \midrule
    \textbf{CR}$\uparrow$ & 2.79 & 2.51 & 1.60 & \cellcolor{bestyellow}4.43 \\
    \textbf{ID}$\uparrow$ & 3.55 & 1.85 & 1.76 & \cellcolor{bestyellow}4.33 \\
    \bottomrule
  \end{tabular}
\end{table}

\vspace{-1em}
\paragraph{LLM-Human Alignment.}
To assess whether the LLM judge can serve as a reliable proxy for human evaluation, we compute the correlation between LLM judge scores and averaged human scores across all 36 paired items.
Content Richness shows strong alignment (Pearson $r = 0.843$, Spearman $\rho = 0.835$), and Interaction Quality likewise (Pearson $r = 0.870$, Spearman $\rho = 0.796$). These results suggest that the LLM-as-Judge framework provides a reliable automated proxy for human assessment on these dimensions.

\subsection{User Study}
\label{sec:userstudy}

We conducted a user study to evaluate \system as an
end-to-end authoring tool, combining a free-form usage
session with a follow-up semi-structured interview.

\subsubsection{System Use}
\label{sec:userstudy_use}

\paragraph{Participants and Procedure.}
We recruited 12 participants (P1--P12) with backgrounds in
visualization, educational technology, or computer science.
After a 10-minute introduction, each participant used
\system to generate two interactive documents on
self-chosen topics, experiencing the full workflow from
topic input through DocSpec editing, style customization,
and generation. Participants then rated the system on
9 items using a 5-point Likert scale, covering usability
(Q1--Q2), controllability (Q3--Q5), output quality
(Q6--Q8), and intent to reuse (Q9).

\paragraph{Results.}
Table~\ref{tab:userstudy} summarizes the ratings.
All items received mean scores above 4.0 on the 5-point scale,
with Usability items (Q1--Q2) achieving a perfect 5.00,
indicating that the system is easy to learn and intuitive to use.
For Controllability, DocSpec (Q3: $4.50 \pm 0.76$) and
chat-based editing (Q5: $4.67 \pm 0.47$) were both rated
highly, while the Style Palette received a comparatively
lower score (Q4: $4.17 \pm 0.49$).
Some participants noted that the lack of inline previews
made it difficult to anticipate how style choices would
affect the final output, a point we revisit in the
qualitative feedback below.
Output quality was rated highly across all dimensions
(Q6--Q8 $\geq 4.58$), with text content and visualizations
both considered informative and engaging.
The strong intent-to-reuse score (Q9: $4.75 \pm 0.43$)
further suggests overall satisfaction with the system.

\begin{table}[t]
  \caption{User study ratings (5-point Likert scale, $n=12$).}
  \label{tab:userstudy}
  \centering
  \small
  \begin{tabular}{c l c}
    \toprule
    & \textbf{Item} & \textbf{Mean $\pm$ SD} \\
    \midrule
    \multicolumn{3}{l}{\textit{Usability}} \\
    Q1 & Easy to learn and use & $5.00 \pm 0.00$ \\
    Q2 & Interface intuitive and satisfying & $5.00 \pm 0.00$ \\
    \midrule
    \multicolumn{3}{l}{\textit{Controllability}} \\
    Q3 & DocSpec gives meaningful control & $4.50 \pm 0.76$ \\
    Q4 & Style Palette gives meaningful control & $4.17 \pm 0.49$ \\
    Q5 & Chat editing refines output effectively & $4.67 \pm 0.47$ \\
    \midrule
    \multicolumn{3}{l}{\textit{Output Quality}} \\
    Q6 & Text content informative and well-written & $4.75 \pm 0.43$ \\
    Q7 & Visualizations engaging and functional & $4.67 \pm 0.47$ \\
    Q8 & Overall document satisfying & $4.58 \pm 0.49$ \\
    \midrule
    \multicolumn{3}{l}{\textit{Intent to Reuse}} \\
    Q9 & Would use again & $4.75 \pm 0.43$ \\
    \bottomrule
  \end{tabular}
\end{table}

\paragraph{Interview.}
Following the system-use session, we conducted a brief
semi-structured interview with each participant to gather
qualitative feedback.
Participants highlighted three main strengths:
(1) DocSpec provides transparent, fine-grained control over
document structure and interaction design \textit{``without
writing any code''} (P4);
(2) the Style Palette and chat-based editing work in a
complementary fashion, with the former for setting the overall
tone and interaction style upfront and the latter for targeted
post-generation refinements; and
(3) the Topic $\to$ Spec $\to$ Style $\to$ Doc workflow felt
natural, with each stage having a clear and distinct role.
Two suggestions for improvement were raised: adding inline
previews within the Style Palette so that style options are
easier to evaluate before generation, and supporting
retrieval-augmented generation for topics requiring specialized
or up-to-date knowledge (e.g., recent research papers or live
datasets). We leave the exploration of these features to
future work.

\section{Conclusion}
We presented \system, a multi-agent framework for controllable interactive document generation. It supports meaningful human collaboration through structured planning (DocSpec), stylistic customization, and chat-based editing. We also introduced \textsc{ViviBench}, a comprehensive evaluation benchmark. Extensive experiments and user studies confirm that \system significantly outperforms existing baselines, offering a highly effective and intuitive authoring experience.
We hope this work lays a foundation for further research
on human-agent collaboration in interactive content
authoring.

\clearpage
\bibliographystyle{ACM-Reference-Format}
\bibliography{custom}

\newpage
\appendix
\section{Ethical Considerations}
The dataset of 101 interactive documents was collected and utilized in strict compliance with applicable copyright regulations. To respect potential copyright concerns, we will only release the URLs of the collected documents rather than distributing their content. For our user study, we obtained explicit informed consent from all participants and rigorously anonymized all interview records to protect personal privacy. Although \system employs LLMs to assist in generation, we mitigate potential risks of AI hallucinations and uncontrollable outputs through a human-in-the-loop paradigm. By allowing users to review and edit the Document Specification (DocSpec) prior to code synthesis, customize styles via the Style Palette, and refine output through chat-based editing, the system ensures that creators retain full control over the authoring intent.

\begin{table}[htbp]
  \caption{Distribution of topics in \textsc{ViviBench} by subject area.}
  \label{tab:dataset}
  \begin{tabular}{lr}
    \toprule
    Subject Area & \# Topics \\
    \midrule
    Algorithms & 25 \\
    Mathematics & 24 \\
    Tools \& Resources & 13 \\
    Physics & 10 \\
    Science & 9 \\
    Other & 6 \\
    Explorable Explanations & 5 \\
    Systems \& Thought Experiments & 4 \\
    Psychology & 2 \\
    Creativity & 2 \\
    Books \& Essays & 1 \\
    \midrule
    \textbf{Total} & \textbf{101} \\
    \bottomrule
  \end{tabular}
\end{table}

\section{Interaction Taxonomy}
\label{app:taxonomy}

From the 101 collected documents, we extracted 482 interaction instances. Three visualization experts collaboratively classified these instances into 8 types based on interaction intent, inspired by Munzner's What-Why-How framework~\cite{munzner2014visualization}:

\begin{itemize}[nosep,leftmargin=*]
  \item \textbf{State Switching} (181, 37.6\%): Switching between discrete options, such as selecting a dataset, algorithm, or display mode.
  \item \textbf{Parameter Exploration} (121, 25.1\%): Adjusting continuous parameters to observe changes, such as tuning a slider for radius, frequency, or threshold.
  \item \textbf{Freeform Construction} (53, 11.0\%): Freely creating content, such as drawing shapes, writing code, editing values, or uploading files.
  \item \textbf{Direct Manipulation} (45, 9.3\%): Dragging objects within a visualization, such as data points, control points, or graph nodes.
  \item \textbf{Temporal Control} (32, 6.6\%): Controlling the time dimension, such as play/pause, stepping, speed adjustment, or timeline scrubbing.
  \item \textbf{Inspection} (24, 5.0\%): Exploring details on demand, such as hover tooltips or cursor tracking.
  \item \textbf{Spatial Navigation} (24, 5.0\%): Navigating in space, such as zooming, panning, or rotating 3D views.
  \item \textbf{Scroll-driven Narrative} (2, 0.4\%): Scrolling drives the narrative progression.
\end{itemize}

\section{SRTC Specifications for Case Study Examples}
\label{app:specs}

The following structured SRTC (State, Render, Transition, Constraint) specifications correspond to the eight interactive visualizations shown in Figure~\ref{fig:cases}. Each specification was produced by the \system Planner agent and served as the code-generation contract for the Executor.

\subsection*{(a) Parameter Exploration — The Lorenz Attractor}
\begin{table}[H]
  \small
  \centering
  \begin{tabularx}{\columnwidth}{@{}lX@{}}
\toprule
\textbf{S} & \texttt{sigma}: slider $[1, 30]$, default 10; \texttt{rho}: slider $[10, 60]$, default 28; \texttt{beta}: constant 2.667; \texttt{trajectory}: derived numerical integration of $dx/dt=\sigma(y-x), dy/dt=x(\rho-z)-y, dz/dt=xy-\beta z$ \\
\midrule
\textbf{R} & \textbullet~A continuously growing 3D phase-space trajectory projected onto a 2D canvas \newline \textbullet~The trail fades over time to emphasize recent motion \newline \textbullet~A slow auto-rotation of the view around the vertical axis \newline \textbullet~Two sliders for $\sigma$ and $\rho$ displayed below the canvas \\
\midrule
\textbf{T} & \textbullet~Adjusting the $\sigma$ slider resets the trajectory and restarts integration from the same initial point \newline \textbullet~Adjusting the $\rho$ slider resets the trajectory, causing the attractor shape to deform or collapse into a stable orbit \\
\midrule
\textbf{C} & For classical values ($\sigma=10$, $\rho=28$, $\beta=8/3$), the trajectory never repeats and draws a distinctive butterfly-shape---demonstrating sensitive dependence on initial conditions. \\
\bottomrule
  \end{tabularx}
\end{table}

\subsection*{(b) Direct Manipulation — Geometric Optics Ray Tracing}
\begin{table}[H]
  \small
  \centering
  \begin{tabularx}{\columnwidth}{@{}lX@{}}
\toprule
\textbf{S} & \texttt{object\_x}: drag-x $[20, \text{lens\_x} - 10]$; \texttt{object\_y}: drag-y $[0, \text{canvas\_height}]$; \texttt{f}: drag-x $[40, 300]$; \texttt{u}: derived $\text{lens\_x} - \text{object\_x}$; \texttt{v}: derived $(u \cdot f) / (u - f)$; \texttt{M}: derived $v / u$; \texttt{image\_type}: derived if $v>0$: Real Inverted, if $v<0$: Virtual Upright, if $v=\infty$: Undefined \\
\midrule
\textbf{R} & \textbullet~A central convex lens whose thickness scales with focal length \newline \textbullet~An orange draggable object arrow on the left of the lens \newline \textbullet~Two green draggable focal points $F$ and $F'$ on the optical axis \newline \textbullet~Three principal rays drawn from the object tip through the lens \newline \textbullet~A colored image arrow on the right (green for real, blue for virtual) \newline \textbullet~An instrument dashboard showing live values of $f, u, v$, and $M$ \newline \textbullet~A status tag indicating image type and orientation \\
\midrule
\textbf{T} & \textbullet~Dragging the orange object arrow horizontally changes $u$ and updates all derived optical values \newline \textbullet~Dragging the object arrow vertically changes the object height and redraws the ray diagram \newline \textbullet~Dragging a focal point ($F$ or $F'$) changes $f$, reshaping both the lens and all derived values simultaneously \\
\midrule
\textbf{C} & The thin lens equation $1/u + 1/v = 1/f$ is always satisfied. When $u < f$, the image distance $v$ becomes negative (virtual image). When $u = f$, the image forms at infinity. \\
\bottomrule
  \end{tabularx}
\end{table}

\subsection*{(c) Inspection — Voronoi Tessellation}
\begin{table}[H]
  \small
  \centering
  \begin{tabularx}{\columnwidth}{@{}lX@{}}
\toprule
\textbf{S} & \texttt{seeds}: 15 points initialized at random positions, moving with slow random velocity; \texttt{mouse\_pos}: hover; \texttt{nearest}: derived $\arg\min_k d(\text{mouse\_pos}, \text{seeds}[k])$; \texttt{min\_dist}: derived $d(\text{mouse\_pos}, \text{seeds}[\text{nearest}])$; \texttt{cell\_region}: derived all pixels closest to \texttt{seeds[nearest]} within 150px from \texttt{mouse\_pos} \\
\midrule
\textbf{R} & \textbullet~An animated canvas of 15 slowly drifting seed points on a dark background \newline \textbullet~On hover: the hovered Voronoi cell illuminated with a purple radial gradient \newline \textbullet~On hover: a dashed gold line connecting the cursor to its nearest seed \newline \textbullet~On hover: a transparent circle of radius \texttt{min\_dist} visualizing the nearest-neighbor envelope \newline \textbullet~The nearest seed highlighted in gold; all others remain purple \\
\midrule
\textbf{T} & \textbullet~Moving the mouse over the canvas updates \texttt{mouse\_pos} continuously \newline \textbullet~Each frame recalculates the nearest seed and updates the illuminated cell, connecting line, and envelope circle in real-time \\
\midrule
\textbf{C} & The illuminated region always corresponds exactly to the Voronoi cell of the nearest seed. Every point within the highlighted region is provably closer to that seed than to any other---demonstrating the core definition of Voronoi partitioning. \\
\bottomrule
  \end{tabularx}
\end{table}

\subsection*{(d) Freeform Construction — Neural Network Forward Propagation}
\begin{table}[H]
  \small
  \centering
  \begin{tabularx}{\columnwidth}{@{}lX@{}}
\toprule
\textbf{S} & \texttt{hidden\_nodes}: click-to-place list of $\{x, y\}$, default $[]$; \texttt{weights}: derived random initialization; \texttt{activations}: derived forward pass (sigmoid); \texttt{output\_val}: derived average of output neuron activations \\
\midrule
\textbf{R} & \textbullet~Three fixed red input nodes ($x_1, x_2, x_3$) on the left \newline \textbullet~Two fixed blue output nodes ($y_1, y_2$) on the right \newline \textbullet~User-placed yellow hidden nodes in the central region \newline \textbullet~Directed arrows connecting all layers; animate green (input$\to$hidden) then orange (hidden$\to$output) during forward pass \newline \textbullet~Activation values displayed on each node \newline \textbullet~An output activation progress bar and percentage readout \newline \textbullet~A 'Send Signal' button and a 'Clear' button \\
\midrule
\textbf{T} & \textbullet~Clicking in the central canvas zone places a new hidden neuron and immediately triggers an animated forward pass \newline \textbullet~Pressing 'Send Signal' replays the forward pass with newly randomized weights \newline \textbullet~Pressing 'Clear' removes all hidden nodes and resets the output bar to 50\% \\
\midrule
\textbf{C} & Adding more hidden neurons generally introduces more non-linearity. The output activation always falls in $(0, 1)$ due to the sigmoid function, regardless of network topology. \\
\bottomrule
  \end{tabularx}
\end{table}

\subsection*{(e) Scroll-driven Narrative — Thermodynamic Entropy}
\begin{table}[H]
  \small
  \centering
  \begin{tabularx}{\columnwidth}{@{}lX@{}}
\toprule
\textbf{S} & \texttt{scroll\_progress}: scroll-wheel $[0, 1]$, default 0; \texttt{partition\_y}: derived $\text{scroll\_progress} \times \text{canvas\_height}$; \texttt{particles}: 150 red + 150 blue bouncing particles; \texttt{entropy\_S}: derived fraction of particles that have crossed to the other side times 100 \\
\midrule
\textbf{R} & \textbullet~A dark split-canvas with 300 bouncing particles: red on left, blue on right \newline \textbullet~A central vertical wall separating the two halves, present only from \texttt{partition\_y} downward \newline \textbullet~A live entropy counter ($S = \dots$) displayed in red in the top-left corner \newline \textbullet~A scroll-progress fill bar on the left edge \\
\midrule
\textbf{T} & \textbullet~Scrolling downward increases \texttt{scroll\_progress}, raising \texttt{partition\_y} and shortening the wall from the top, allowing particles to mix \newline \textbullet~Scrolling upward decreases \texttt{scroll\_progress}, lowering \texttt{partition\_y} and re-blocking particle passage \\
\midrule
\textbf{C} & As the wall is removed ($\text{scroll\_progress} \to 1$), particles irreversibly mix---entropy $S$ increases monotonically. The constraint $\Delta S \ge 0$ maps directly to the unidirectional scroll interaction, visualizing time's arrow. \\
\bottomrule
  \end{tabularx}
\end{table}

\subsection*{(f) Spatial Navigation — The Möbius Strip}
\begin{table}[H]
  \small
  \centering
  \begin{tabularx}{\columnwidth}{@{}lX@{}}
\toprule
\textbf{S} & \texttt{rotX}: drag-y $[-3.14, 3.14]$, default 0.5; \texttt{rotY}: drag-x $[-3.14, 3.14]$, default -0.5; \texttt{surface}: derived parametric mesh $x(u,v)=(R+v\cos(u/2))\cos(u), y(u,v)=v\sin(u/2), z(u,v)=(R+v\cos(u/2))\sin(u)$ \\
\midrule
\textbf{R} & \textbullet~A 3D polygon mesh of the Möbius strip rendered using the Painter's Algorithm \newline \textbullet~Faces colored with a teal-to-sky-blue gradient mapped to the $u$ parameter \newline \textbullet~Depth shading simulating a diffuse light source \newline \textbullet~Numeric overlays showing the current \texttt{rotX} and \texttt{rotY} viewing angles \\
\midrule
\textbf{T} & \textbullet~Clicking and dragging horizontally on the canvas updates \texttt{rotY}, rotating the strip around the vertical axis \newline \textbullet~Clicking and dragging vertically updates \texttt{rotX}, tilting the strip forward or backward \newline \textbullet~Releasing the mouse locks the current orientation \\
\midrule
\textbf{C} & No matter how the strip is rotated, a continuous path along its surface always returns to the starting point in a mirrored orientation, demonstrating the strip's single-sidedness. \\
\bottomrule
  \end{tabularx}
\end{table}

\subsection*{(g) State Switching — Quantum Electron Orbitals}
\begin{table}[H]
  \small
  \centering
  \begin{tabularx}{\columnwidth}{@{}lX@{}}
\toprule
\textbf{S} & \texttt{orbital\_state}: segmented-button [\texttt{1s}, \texttt{2p}, \texttt{3d}], default \texttt{1s}; \texttt{wavefunction}: derived $\psi(r,\theta) = \text{radial\_part}(n,l) \times \text{angular\_part}(l,m)$; \texttt{density\_cloud}: derived Monte Carlo sampling, accept point $(x,y)$ with probability $\propto |\psi(x,y)|^2$ \\
\midrule
\textbf{R} & \textbullet~A 2D canvas with coordinate axes and a nucleus dot at the origin \newline \textbullet~Points progressively sampled and plotted, accumulating into a probability density cloud \newline \textbullet~A wavefunction equation display that updates to reflect the active orbital state \newline \textbullet~A segmented button control with options \texttt{1s}, \texttt{2p}, and \texttt{3d} \\
\midrule
\textbf{T} & \textbullet~Clicking a segment button sets \texttt{orbital\_state} to the selected orbital \newline \textbullet~Switching state clears all existing sample points and triggers a new Monte Carlo sampling run \\
\midrule
\textbf{C} & Each orbital state produces a distinct, characteristic spatial density pattern: \texttt{1s} is spherically symmetric, \texttt{2p} is a dumbbell shape with a nodal plane, and \texttt{3d} forms a four-lobed clover pattern---matching theoretical predictions. \\
\bottomrule
  \end{tabularx}
\end{table}

\subsection*{(h) Temporal Control — Fourier Series Epicycles}
\begin{table}[H]
  \small
  \centering
  \begin{tabularx}{\columnwidth}{@{}lX@{}}
\toprule
\textbf{S} & \texttt{time}: playback $[0, \infty]$, default 0; \texttt{is\_playing}: toggle, default true; \texttt{n\_harmonics}: slider $[1, 15]$ (step 2), default 5; \texttt{wave}: derived last 500 $y$-values of the epicycle tip \\
\midrule
\textbf{R} & \textbullet~A chain of rotating circles (epicycles), each at a frequency proportional to its harmonic index \newline \textbullet~A dot tracing the tip of the outermost epicycle \newline \textbullet~A reconstructed square wave drawn on the right by recording the tip's $y$-position over time \newline \textbullet~A dashed line connecting the epicycle tip to the leading edge of the wave \newline \textbullet~A Play/Pause button and a harmonic count slider \\
\midrule
\textbf{T} & \textbullet~Pressing Play/Pause starts or freezes the rotation of all epicycles \newline \textbullet~Dragging the harmonic slider adds or removes outer epicycles in odd increments, immediately reshaping the output wave \\
\midrule
\textbf{C} & As \texttt{n\_harmonics} increases toward infinity, the reconstructed wave converges to a perfect square wave, demonstrating Fourier's theorem that any periodic signal decomposes into sinusoids. \\
\bottomrule
  \end{tabularx}
\end{table}

\end{document}